# Scientific capabilities and advantages of the 3.6 meter optical telescope at Devasthal, Uttarakhand

**Amitesh Omar[1],\*, Brijesh Kumar[1], Maheswar Gopinathan[1] and Ram Sagar[1,2]**

[1]Aryabhatta Research Institute of Observational Sciences, Manora Peak, Nainital 263 002, India
[2]Indian Institute of Astrophysics, 2nd Block, Koramangala, Bengaluru 560 034, India

**India's largest 3.6 m aperture optical telescope has been successfully installed in the central Himalayan region at Devasthal, Nainital district, Uttarakhand. The primary mirror of the telescope uses the active optics technology. The back-end instruments, enabling spectroscopic and photometric imaging of the celestial sky are designed and developed by ARIES along with other Indian institutes. The Devasthal optical telescope in synergy with two other highly sensitive telescopes in the country, namely GMRT operating in the radio wavebands and AstroSat operating in the high-energy X-ray, ultraviolet and visual wavebands, will enable Indian astronomers to carry out scientific studies in several challenging areas of astronomy and astrophysics.**

## Introduction

THE largest optical telescope in India until recently was the 2.3-m Vainu Bappu Telescope in Kavalur[1]. Other major optical telescopes in the country have diameters between ~1 and 2 m (ref. 2). The Indian astronomers planned to install a modern 4-m class optical telescope in the country many years ago. The main reasons for this were to supplement the observations taken at other wavelengths with the Indian Giant Meterwave Radio Telescope (GMRT) and the Indian space observatory AstroSat, as well as to fill the large longitudinal gap in the Asia region between western Russia (Caucasus mountains at longitude of 41°E) and Australia (Siding Spring mountains at longitude of 149°E). This requirement has now been fulfilled with the installation of the 3.6-m Devasthal Optical Telescope (DOT) in the central Himalayan region of Uttarakhand[3]. It is optimized to operate in the visible and near infrared wavelengths up to ~3 μm. DOT, an Indo-Belgian venture, established by ARIES, Nainital was technically activated jointly by the Prime Ministers of India and Belgium on 30 March 2016 in the presence of the Minister of Science and Technology, Government of India.

DOT is a technologically advanced moderate aperture telescope designed and built by Advanced Mechanical and Optical System, Belgium[4]. The telescope uses two hyperbolic mirrors making the Ritchey–Crétien optical system, which provides images free from major optical aberrations over a reasonable field of view. The effective focal length of the telescope is ~32,400 mm, providing an image scale of ~6.4 arcsec/mm. The primary mirror weighing ~4300 kg is supported on 69 precisely controlled push-only axial actuators. This configuration, known as the active optics system, controls the shape of the primary mirror at all elevations and provides the best focus for images of celestial objects. The first aluminization of the primary mirror to make it highly reflective (~90%) was carried out at Devasthal in 2015 using a magnetron vacuum chamber designed and developed in the country[5]. The telescope is housed in the cylindrical dome of total height of ~32 m. The telescope building is made mostly of steel to enable rapid cooling in the evening and hence minimizing the thermal gradients in the air inside the building. Several exhaust fans in the dome also help in rapidly attaining thermal equilibrium inside the building with the outside air in the evening, before the start of observations. These features of the building help in obtaining the best dome seeing and hence sharp images of the celestial sky. Figure 1 shows photographs of the telescope and the building. Some binary stars could be resolved to an angular separation of ~0.4 arcsec, during the test observations with DOT carried out in December 2015. Figure 2 shows the images of the binary stars obtained using DOT. The seeing value recorded in these images is at par with other major international optical telescope sites.

The Devasthal site at a peak altitude of ~2450 m amsl was chosen for the 3.6-m telescope after detailed atmospheric characterization carried out between 1990 and 2000 (ref. 6). Devasthal is easily approachable by road in about 2 h from the nearest railway terminal at Kathgodam at a distance of nearly 55 km. The infrastructural and weather conditions at Devasthal are hospitable round the year. The night sky is considerably dark due to the thin population in this remote hilly area near Nainital. The wind conditions remain low to moderate most of the time. The diurnal temperature variations are minimal and gradual. A total of about 200 photometric, cloud-free usable nights per year are expected. In the best observing atmospheric conditions found between November and January the relative humidity frequently drops below 20%, making the site conducive for both visual and near infrared band observations.

*For correspondence. (e-mail: aomar@arises.res.in)

Unprecedented progress has been made during the past two decades in several areas of astronomy and astrophysics. Both large and small–moderate aperture optical telescopes have played important roles in this progress. Small-aperture optical telescopes have limited sensitivity particularly in carrying out deeper spectroscopic observations of celestial objects. A large-aperture telescope is required, as the entire light needs to be divided into finer bandwidths of a fraction of an Angstrom in high-resolution spectroscopic observations. The sensitive spectroscopic observations are required to detect particular features such as Balmer lines of hydrogen and transition lines of various elements (oxygen, nitrogen, calcium, iron, etc.) associated with stars and interstellar medium. The detections of spectral lines of various elements at high resolution enable astronomers to probe and understand stellar nuclear synthesis, patterns of elemental enrichment histories, formation epochs and dynamics and precise distances of the celestial objects.

The present emphasis worldwide in astronomy and astrophysics is to deeply probe and understand the life cycle of stars, occurrences of planets around stars, variable and active stars, formation of galaxies at high redshifts, elusive objects such as black holes, and interstellar and cosmological physics and chemistry associated with gas, dust, dark matter and magnetic fields in the Universe. An access to at least a moderate-sized optical telescope is required to carry out scientific studies in all of these areas. The research priorities and interests of the Indian astronomy community are wide and cover areas right from the solar system astronomy to extra-solar planets, and galactic astronomy to high redshift cosmological science. The 3.6-m DOT dedicated to Indian (93% of the total time) and Belgian (7% of the total time) astronomers will make it possible to carry out scientific studies in the frontline areas of astronomy and astrophysics, particularly those which require spectroscopic and deep photometric observations of faint celestial objects.

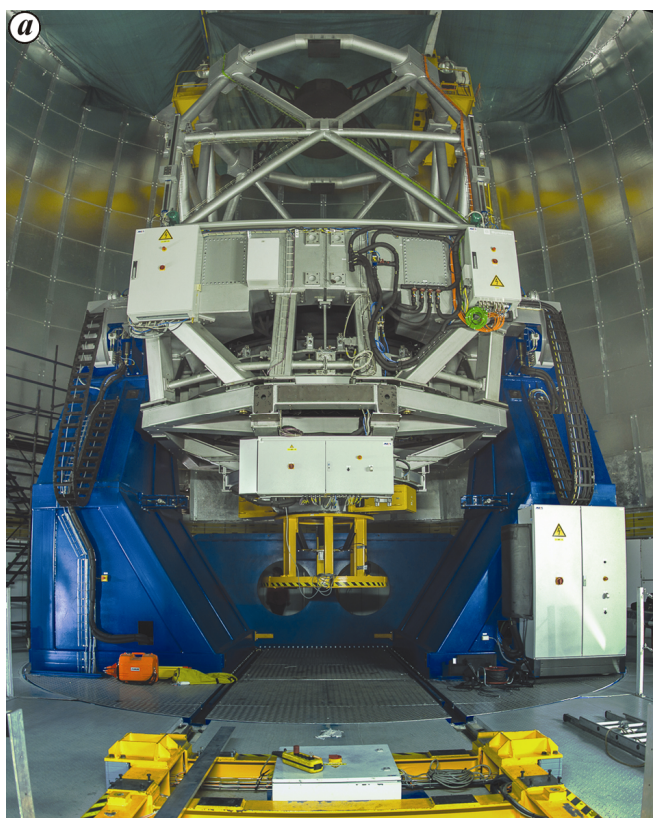

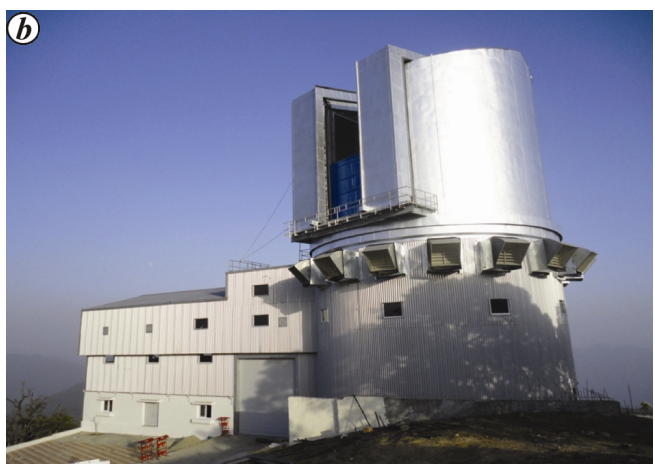

**Figure 1.** *a*, Photograph of the 3.6-m Devasthal Optical Telescope (DOT); *b*, Photograph of the building housing the 3.6-m DOT.

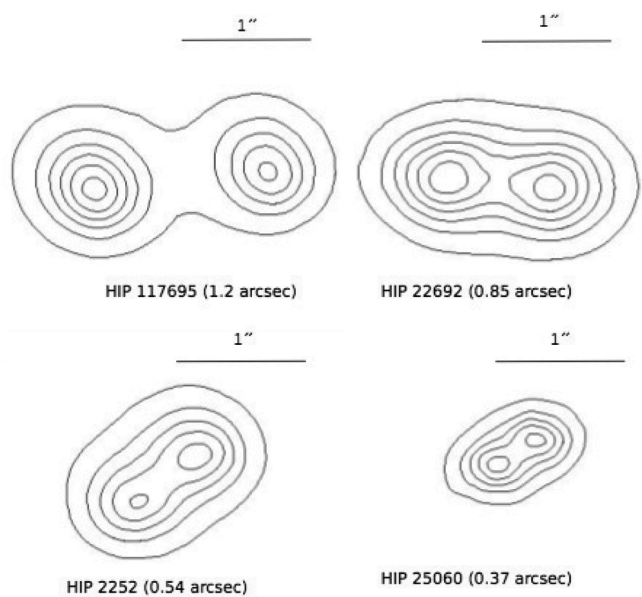

**Figure 2.** The iso-intensity contour images of close-separation binary stars observed with the 3.6-m DOT. Stars up to 0.4″ angular separation were resolved in the best seeing condition. This seeing is at par with those at major ground-based international large telescope sites.

Scientific areas where DOT is expected to play an important role include asteroseismology, fast milli-second variability of stars and black-hole candidates, detection of extra-solar planets and cool dwarf stars, detection of young galaxies at high redshifts, and optical follow-up observations of the most energetic catastrophic events in the Universe such as gamma ray bursts (GRB) and supernovae. The study of variability helps in understanding internal structure and physical conditions of stars, which are otherwise not possible to probe directly. The rapid optical variability of X-ray emitting black-hole and neutron star candidates helps in understanding accretion and mass flow phenomena in these most compact objects known in the Universe. Many of these transient-class observations falling in time-critical category require observations with a telescope as quickly or as often as possible, as they either tend to fade rapidly or emit semi-periodic variable patterns of emissions from time to time. Such observations require telescopes of sufficient aperture to be available in the night period at the time of occurrence or expectation of the event. The geographical latitude advantages of the Indian subcontinent are well established for such time-critical transient events as the 1-m telescope at Nainital played an important role during the early optical follow-up observations of the first few recorded GRB afterglows[7].

The aperture of DOT is most suitable for observations where photon count rates are feeble such as in medium to high resolution spectroscopy, observations of faint and distant sources, and fast milli-second observations of variable sources. For example, with a total collecting area of $\sim 10^5$ sq. cm with the 3.6-m aperture, and assuming an overall system (telescope + instrument) efficiency of 30%, for a very faint source at the visual band (yellow-green) magnitude of 25, about 3 photons/sec will be registered. For a comparison, the human eye has the faintest detection limit at about 7 magnitude that is nearly 10 million times brighter than a source at $V \sim 25$ magnitude, which should be possible to detect with DOT using sensitive charge coupled device (CCD) cameras. A source up to $V \sim 22$ magnitude should be detectable in spectroscopy mode, after dividing the visible light from the source into 2000 wavelength resolution elements. For rapid variability studies, DOT along with an electron-multiplying CCD camera should enable observations of sources down to ~18 magnitude at ~10 milli-second cadence.

Considering the importance of DOT as the largest optical–near infrared aperture in the country, several Indian institutes, including ARIES planned to develop cutting-edge back-end instruments for this telescope. Scientists and engineers from ARIES are actively participating in the design and development of these back-end instruments. The following back-end instruments are being developed for DOT:

(i) A 4096 × 4096 (~16 million) pixels (~61 mm chip) back-illuminated liquid nitrogen-cooled CCD imager with broadband color filters, covering a field of view of ~6.4 arcmin for deep photometry observations has been developed by ARIES[3].

(ii) A low-dispersion visible wavelength spectroscopy-cum-imager instrument, similar to the versatile Faint Object Spectrograph and Camera (FOSC) instruments, has also been developed by ARIES[8]. The optics and mechanical parts of this spectrograph were designed and assembled within the country. The precision mechanical parts were manufactured in small-scale industries near Delhi, while the custom-made optical elements were imported. The spectrograph contains 13 optical lenses, optical colour filters, dispersive elements such as prisms and grisms providing a spectral resolution up to 2000. The spectrograph will provide a field of view of ~14 arcmin in the sky after converting the $f/9$ optical beam of the telescope into a faster $\sim f/4.2$ beam. This instrument is the first of its kind developed within the country.

(iii) Eight fibre-based integral field spectrographs are being designed and developed at the Inter University Center for Astronomy and Astrophysics, Pune in collaboration with ARIES, and scientists from South Korea[9]. This highly complex and technologically advanced set of spectrographs, being developed for the first time in the country, will have 16 field-deployable fibre bundles each having 144 fibres and covering a field of view of ~64 arcsec$^2$ in the sky. The individual fibres will provide a spatial resolution of ~0.8 arcsec. This instrument will enable astronomers to carry out detailed spectroscopic observations over a two-dimensional field, thereby making it most suitable for observations of diffuse objects such as galaxies. The limiting sensitivity of this instrument is expected to be at $\sim 10^{-17}$ erg/s/cm$^2$/arcsec$^2$ with a resolution of nearly 4 Å.

(iv) With joint efforts made by Tata Institute of Fundamental Research (TIFR), Mumbai and ARIES, a near-infrared imager-cum-spectrograph is being built by MKIR (Mouna Kea Infrared) in USA. This instrument will be capable of observing in the wavelength range 0.6–2.5 μm at spectral resolution below 3000. The main detector uses the Hawaii-2RG near-infrared focal-plane array. This spectrometer is expected to detect sources at K-band magnitude of ~16 in an hour of exposure time. A near-infrared imager developed by TIFR sensitive in the wavelength range of 1–3.7 μm is also possible to use on DOT.

Apart from the above-mentioned instruments, development of a dedicated three-channel fast photometer, a high spectral resolution Echelle spectrometer and a spectro-polarimeter are under discussion and in planning stages. A high-resolution spectrograph is helpful in precise determination of elemental abundances in stars and interstellar medium, chemical enrichment history of the Milky-way and to detect planets around other stars. The polarimeter will enable astronomers to probe magnetic fields and interstellar dust.

The 3.6-m DOT enjoying its geographical location advantage with a fleet of upcoming sensitive back-end instruments with photometric and spectroscopic observation capabilities in the visible and near infrared wavelength range can make it one of the most productive and desired telescopes by the national and international community. The prime aim of DOT will be to strengthen India's capabilities in astronomy and astrophysics. Being at an easily accessible location, DOT is suitable for astronomers to visit and make a direct use of the telescope. This is particularly beneficial for young students, scientists and engineers, who can get trained through practical experience of observing and working with a telescope. The international collaborations are expected to get strengthened in the area of astronomy and astrophysics, through the use of this facility by the international community in collaboration with Indian astronomers. The Devasthal observatory can greatly benefit by joining certain global observing programmes. The existing Belgo-Indian Network for Astronomy & Astrophysics (BINA) can play an important role in maximizing scientific output from DOT. It will be beneficial for the observatory to take up certain challenging developmental activities such as adaptive optics and speckle interferometry, and design and assembly of next-generation back-end instruments for the telescope.

India has already established two major astronomy observatories, namely GMRT, for observations at radio wavelengths, and India's first multi-wavelength space astronomy telescope AstroSat, with observing capabilities in high energy X-ray, UV/visual bands. With the addition of DOT as the most sensitive optical telescope in the country, Indian astronomers will be able to carry out sensitive observations of the celestial sky over a large span of the electromagnetic spectrum in X-rays, ultraviolet, visible, near infrared and radio wavelengths using the facilities within the country. It is expected that coordinated observations using DOT, AstroSat and GMRT will be actively pursued. DOT, a 4-m class optical observing facility of national importance, with the excellent atmospheric conditions at Devasthal has immense potential to make significant impact in the international astronomy community in coming years.

ACKNOWLEDGEMENTS. We acknowledge academic, technical, and administrative supports received from the staff of ARIES. The support received from scientists and engineers from various institutes and organizations is also acknowledged. The dedicated and timely support received from the Department of Science and Technology (DST), Government of India, ARIES Governing Council and Project Management Board is duly acknowledged. We thank the officiating director Dr Wahab Uddin and previous project managers, Dr B. B. Sanwal and Dr A. K. Pandey for their contributions towards the project. The tireless efforts made by Prof. G. Srinivasan to guide the ARIES team during completion of the enclosure and installation of the telescope are acknowledged. R.S. thanks the National Academy of Sciences, India for providing financial assistance at his present work place. We thank the editor of *Current Science* for an invitation to write this article. All authors contributed equally to this work.

doi: 10.18520/cs/v113/i04/682-685